\documentclass[preprintnumbers,amsmath,amssymb,showpacs,twocolumn]{revtex4}

\usepackage{graphicx}
\usepackage{dcolumn}
\usepackage{bm}

\begin{document}

\title{ Two-step deterministic remote preparation of an arbitrary quantum state in the whole Hilbert space}

\author{Meiyu Wang, Fengli Yan}
\thanks{Corresponding author. flyan@mail.hebtu.edu.cn}
\affiliation { College of Physics  Science and Information
Engineering, Hebei Normal University, Shijiazhuang 050016, China
\\
Hebei Advanced Thin Films Laboratory,  Shijiazhuang 050016, China}

\date{\today}

\begin{abstract}

 We present a two-step exact remote state preparation  protocol
 of an arbitrary qubit with the aid of a three-particle Greenberger-Horne-Zeilinger state.
 Generalization of this protocol for higher-dimensional Hilbert
 space systems among three parties is also given. We show that
 only single-particle von Neumann measurement, local operation and
 classical communication are necessary. Moreover, since the overall
 information of the quantum state can be divided into two different
 parts, which may be at different locations, this protocol may be
 useful in the quantum information field.
  \end{abstract}


\maketitle

\section{Introduction}

Quantum information theory has produced many interesting and
important developments that are not possible classically in recent
years, in which quantum entanglement and classical communication are
two elementary resources. Two surprising discoveries in this area
are teleportation and remote state preparation (RSP). Quantum
teleportation process, originally proposed by Bennett et al \cite
{s1}, can transmit an unknown quantum state from a sender (called
Alice) to a spatially distant receiver (called Bob) via a quantum
channel with the help of some classical information. Recently, Lo
\cite {s2}, Pati \cite {s3} and Bennett et al \cite {s4} have
presented an interesting application of quantum entanglement, i.e.,
remote state preparation that correlates closely to teleportation.
RSP is called "teleportation of a known quantum state", which means
Alice knows the precise state that she will transmit to Bob. Her
task is to help Bob construct a state that is unknown to him by
means of a prior shared entanglement and a classical communication
channel. So the goal of RSP is the same as that of quantum
teleportation. The main difference between RSP and teleportation is
that in the former Alice is assumed to know completely the state to
be prepared remotely by Bob; in particular, Alice need not own the
state, but only know information about the state, while in the
latter Alice must own the transmitted state, but neither she nor Bob
has knowledge of the transmitted state.

So far, RSP has attracted much attention \cite
{s5,s6,s7,s8,s9,s10,s11,s12,s13,s14,s20}. There are  many kinds of
RSP methods in theory, such as low-entanglement RSP \cite {s5},
higher-dimension RSP \cite {s6}, optimal RSP \cite {s7}, oblivious
RSP \cite {s8}, RSP without oblivious conditions \cite {s9}, RSP for
multiparties \cite {s10}, and continuous variable RSP in phase
space\cite {s11,s12}, etc. On the other hand, some RSP schemes have
been implemented experimentally with the technique of NMR \cite
{s15} and spontaneous parametric down-conversion \cite {s16,s17}. In
addition, some authors have also investigated the RSP protocol using
different quantum channels such as partial EPR pairs \cite {s18} and
three-particle Greenberger-Horne-Zeilinger (GHZ) state \cite {s19}.
To our best knowledge, up to now there is no RSP protocol which
determinately generate  an arbitrary qubit with unit success
probability. They mainly concentrate on RSP of some special
ensembles of a quantum state. For example, some schemes discuss how
to successfully remotely prepare the state in subspace of the whole
real Hilbert space or chosen from equatorial line on Bloch sphere.

In this paper, we propose a two-step deterministic RSP protocol via
 previously shared entanglement, a single-particle von Neumann
measurement, local operation and classical communication.
Generalization of this protocol for higher-dimensional Hilbert space
systems among three parties is also presented. We will see that the
overall information of an arbitrary quantum state can be divided
into two different parts. They are expressed by $\theta$ and
$\varphi$ respectively, which  may be at different locations. So
this protocol may be useful in the quantum information field, such
as quantum state sharing, converging the split information at one
point, etc.

\section{Deterministic RSP of an arbitrary qubit using a  GHZ state as a quantum channel}

Let us consider a pure  state $|\psi\rangle\in H=C^2$ which is the
state of a qubit. An arbitrary qubit can be represented as
\begin{equation}
|\psi\rangle=\alpha|0\rangle+\beta|1\rangle,
\end{equation}
where we can choose $\alpha$ to be real and $\beta$ to be complex
number and $|\alpha|^2+|\beta|^2=1$. This qubit can be represented
by a point on the unit two-dimensional sphere, known as Bloch
sphere, with the help of two real parameters $\theta$ and $\varphi$.
So we can rewrite  Eq.(1) as
\begin{equation}
|\psi\rangle=\cos
(\theta/2)|0\rangle+\sin(\theta/2)e^{i\varphi}|1\rangle.
\end{equation}
Now Alice wants to transmit the above qubit to Bob. The quantum
channel shared by Alice and Bob is the three-particle GHZ state
\begin{equation}
|\Phi\rangle =\frac {1}{\sqrt 2}(|000\rangle+|111\rangle)_{123}.
\end{equation}
 The particles 1 and 2
belong to Alice and the particle 3 is held by Bob. As a matter of
fact, the state $|\Phi\rangle$ can be easily generated from the Bell
state $\frac {1}{\sqrt 2}(|00\rangle+|11\rangle)_{23}$, because
particles 1 and 2 belong to Alice. A Controlled-Not gate can
transform $\frac {1}{\sqrt
2}|0\rangle_1(|00\rangle+|11\rangle)_{23}$ into $|\Phi\rangle$, when
particle 2 and particle 1 are  a controlled qubit and a target
qubit, respectively. We suppose the qubit $|\psi\rangle$ is known to
Alice, i.e. Alice knows $\theta$ and $\varphi$ completely, but Bob
does not know them at all. Since Alice knows the state she can
choose to measure the particles 1 and 2 in any basis she wants.
First, Alice performs a projective measurement on particle 1. The
measurement basis chosen by Alice is a set of mutually orthogonal
basis vectors $\{|\phi\rangle, |\phi_{\perp}\rangle\}$, which is
related to the computation basis $\{|0\rangle, |1\rangle\}$ in the
following manner
\begin{eqnarray}
&&|\phi\rangle_1=\cos(\theta/2)|0\rangle_1+\sin(\theta/2)|1\rangle_1,\nonumber\\
&&|\phi_{\perp}\rangle_1=\sin(\theta/2)|0\rangle_1-\cos(\theta/2)|1\rangle_1.
\end{eqnarray}
By this change of basis, the normalization and orthogonality
relation between basis vectors are preserved. Using Eq.(4), we can
express Eq.(3) as
\begin{equation}
|\Phi\rangle=\frac {1}{\sqrt
2}(|\phi\rangle_1|\Psi\rangle_{23}+|\phi_{\perp}\rangle_1|\Psi_{\perp}\rangle_{23}),
\end{equation}
where
\begin{eqnarray}
&&|\Psi\rangle_{23}=\cos(\theta/2)|00\rangle_{23}+\sin(\theta/2)|11\rangle_{23},\nonumber\\
&&|\Psi_{\perp}\rangle_{23}=\sin(\theta/2)|00\rangle_{23}-\cos(\theta/2)|11\rangle_{23}.
\end{eqnarray}
Now Alice measures the particle 1. For example, if Alice's von
Neumann measurement result is $|\phi\rangle_1$, then the state of
particles 2 and 3, as shown by Eq.(5), will collapse into
$|\Psi\rangle_{23}$. Next, Alice performs another projective
measurement on particle 2. The measurement basis is also a set of
mutually orthogonal basis vectors $\{|\eta\rangle,
|\eta_{\perp}\rangle\}$, the relation between the measurement basis
$\{|\eta\rangle, |\eta_{\perp}\rangle\}$ and the computation basis
$\{|0\rangle, |1\rangle\}$ is given by
\begin{equation}
|\eta\rangle_2=\frac {1}{\sqrt
2}(|0\rangle_2+e^{-i\varphi}|1\rangle_2), ~~
|\eta_{\perp}\rangle_2=\frac {1}{\sqrt
2}(|0\rangle_2-e^{-i\varphi}|1\rangle_2).
\end{equation}
 Then,
we have
\begin{equation}
|\Psi\rangle_{23}=\frac {1}{\sqrt
2}(|\eta\rangle_2|\psi\rangle_{3}+|\eta_{\perp}\rangle_2|\psi'\rangle_{3}),
\end{equation}
where
\begin{eqnarray}
&&|\psi\rangle=\cos(\theta/2)|0\rangle+\sin(\theta/2)e^{i\varphi}|1\rangle,\nonumber\\
&&|\psi'\rangle=\cos(\theta/2)|0\rangle-\sin(\theta/2)e^{i\varphi}|1\rangle.\end{eqnarray}
If Alice's von Neumann measurement result is $|\eta\rangle_2$, the
particle 3 can be found in the original state $|\psi\rangle$, which
is nothing but the remote state preparation of the known qubit. If
the outcome of Alice's measurement result is
$|\eta_{\perp}\rangle_2$, then the classical communication from
Alice will tell Bob that he has obtained a state $|\psi'\rangle$.
Bob can carry out the unitary operation $\sigma_z=|0\rangle\langle
0|-|1\rangle\langle1|$ on his particle 3. That is
\begin{equation}
\sigma_z|\psi'\rangle=\cos(\theta/2)|0\rangle+\sin(\theta/2)e^{i\varphi}|1\rangle=|\psi\rangle.
\end{equation}
This means after Bob's unitary operation the state $|\psi\rangle$
has already been prepared in Bob's qubit.

Surely it is possible for Alice to get the state
$|\phi_{\perp}\rangle_1$ after her measurement on particle 1. If so,
she will choose another measurement basis
$\{|\xi\rangle,|\xi_{\perp}\rangle\}$ on particle 2, which are
written as
\begin{equation}
|\xi\rangle_2=\frac {1}{\sqrt
2}(|1\rangle_2+e^{-i\varphi}|0\rangle_2), ~~
|\xi_{\perp}\rangle_2=\frac {1}{\sqrt
2}(|1\rangle_2-e^{-i\varphi}|0\rangle_2).
\end{equation}
Obviously, the basis vectors $\{|\xi\rangle, |\xi_{\perp}\rangle\}$
and $\{|\eta\rangle, |\eta_{\perp}\rangle\}$ can be mutually
converted by a unitary operation $\sigma_x=|0\rangle\langle
1|+|1\rangle\langle 0|$.

After Alice's measurement, for each collapsed state Bob can employ
an appropriate unitary operation to convert it to the prepared state
$|\psi\rangle$ except for an overall trivial factor. Here we do not
depict them one by one anymore. As a summary, Bob's corresponding
unitary operations to Alice's measurement results are listed in
Table I. One can easily work out that the total probability of RSP
is 1 though the classical communication cost is 2 bits.
\begin{table}
\caption{Alice's measurement basis on particle 1 (MB1), Alice's
measurement outcome for particle 1 (AMO1), Alice's measurement basis
on particle 2 (MB2), Alice's measurement outcome for particle 2
(AMO2), the collapse states for particle 3 (CS3) and Bob's
appropriate unitary operation (BAUO)} {\footnotesize
\begin{tabular}{|c|c|c|c|c|c|} \hline
 MB1 & AMO1& MB2 & AMO2& CS3 & BAUO\\\hline
 $\{|\phi\rangle,|\phi_{\perp}\rangle\}$ & $|\phi\rangle_1$
 &$\{|\eta\rangle,|\eta_{\perp}\rangle\}$&
 $|\eta\rangle_2$ & $\begin{array}{c}\cos(\theta/2)|0\rangle+\\\sin(\theta/2)e^{i\varphi}|1\rangle\end{array} $ &
 $I$\\\hline
 $\{|\phi\rangle,|\phi_{\perp}\rangle\}$ & $|\phi\rangle_1$
 &$\{|\eta\rangle,|\eta_{\perp}\rangle\}$&
 $|\eta_{\perp}\rangle_2$ & $\begin{array}{c}\cos(\theta/2)|0\rangle-\\\sin(\theta/2)e^{i\varphi}|1\rangle\end{array} $ &
 $\sigma_z$\\\hline
 $\{|\phi\rangle,|\phi_{\perp}\rangle\}$ & $|\phi_{\perp}\rangle_1$
 &$\{|\xi\rangle,|\xi_{\perp}\rangle\}$&
 $|\xi\rangle_2$ & $\begin{array}{c}\sin(\theta/2)e^{i\varphi}|0\rangle-\\\-\cos(\theta/2)|1\rangle\end{array} $ &
 $\sigma_z\sigma_x$\\\hline
 $\{|\phi\rangle,|\phi_{\perp}\rangle\}$ & $|\phi_{\perp}\rangle_1$
 &$\{|\xi\rangle,|\xi_{\perp}\rangle\}$&
 $|\xi_{\perp}\rangle_2$ & $\begin{array}{c}\sin(\theta/2)e^{i\varphi}|0\rangle+\\\cos(\theta/2)|1\rangle\end{array} $ &
 $\sigma_x$\\\hline
\end{tabular}}
\end{table}

By the above analysis, one can easily see that unlike the standard
teleportation of an unknown qubit, here, we do not require a
Bell-state measurement, which is still more difficult according to
the present-day technologies. Only  single-particle von Neumann
measurement and local operation are necessary. On the other hand,
the total probability of RSP for an arbitrary qubit is 1 while in
the previous schemes only the probability of RSP of some special
ensembles of qubit is 1. In addition, what deserves mentioning here
is that in this protocol, the overall information of the qubit,
which is expressed by $\theta$ and $\varphi$, can be divided into
two parts. We must first prepare the part $\theta$ and then prepare
the remainder part $\varphi$, which  can not be transposed. This
indicates that the two parts of information are not equal with each
other.

 As mentioned above, we need only the single-particle
measurement and local operation. So, the particle 1 and 2 may be at
different locations. In this case, $|\Phi\rangle$  is a real GHZ
state. It is natural to generate it to the three-party RSP.

\section{RSP of higher-dimensional quantum state for three parties}

 In
this section, we wish  to generalize the RSP protocol to systems
with larger than two-dimensional  Hilbert space among three parties.

First we consider the case that two parties (Alice and Bob)
collaborate with each other to prepare a 4-dimensional quantum state
at Charlie's location. A  quantum state
\begin{eqnarray}
&|\psi\rangle=&\cos\gamma_1|0\rangle+\sin\gamma_1\cos\gamma_2e^{i\alpha_1}|1\rangle\nonumber\\
&&+\sin\gamma_1\sin\gamma_2\cos\gamma_3e^{i\alpha_2}|2\rangle\nonumber\\
&&+\sin\gamma_1\sin\gamma_2\sin\gamma_3e^{i\alpha_3}|3\rangle
\end{eqnarray}
in a four-dimensional Hilbert space can be parameterized by six
parameters $\gamma_1,\gamma_2,\gamma_3,\alpha_1,\alpha_2$ and
$\alpha_3$ such that $0\leq \gamma_1,\gamma_2,\gamma_3\leq \pi/2$
and $0\leq \alpha_1,\alpha_2,\alpha_3\leq 2\pi$. Alice and Bob know
$\gamma_1,\gamma_2,\gamma_3$ and $\alpha_1,\alpha_2$ and $\alpha_3$
partly respectively, that is, Alice has information of
$\gamma_1,\gamma_2,\gamma_3$, and Bob has information
$\alpha_1,\alpha_2$ and $\alpha_3$. The quantum channel shared by
Alice, Bob and Charlie is a 4-level maximally GHZ state
\begin{equation}
|\Phi\rangle_{ABC}=\frac
{1}{2}(|000\rangle+|111\rangle+|222\rangle+|333\rangle)_{ABC},
\end{equation}
where particle $A$, $B$ and $C$ belong to Alice, Bob and Charlie
respectively. The method is similar to the case of qubit. First
Alice must find a set of orthogonal basis vectors to perform a
generalized projective measurement on particle $A$. We shall see
below, there exist many sets of orthogonal basis vectors that
include the state (12). One such set can be obtained by applying a
specific unitary transformation on the computational basis vectors
\begin{eqnarray}
&U(\gamma_1,\gamma_2,\gamma_3)|0\rangle=|\phi_0\rangle=&\cos\gamma_1|0\rangle+\sin\gamma_1\cos\gamma_2|1\rangle\nonumber\\
&&+\sin\gamma_1\sin\gamma_2\cos\gamma_3|2\rangle\nonumber\\
&&+\sin\gamma_1\sin\gamma_2\sin\gamma_3|3\rangle,\nonumber\\
&U(\gamma_1,\gamma_2,\gamma_3)|1\rangle=|\phi_1\rangle=&-\sin\gamma_1\cos\gamma
_2|0\rangle+\cos\gamma_1|1\rangle\nonumber\\
&&-\sin\gamma_1\sin\gamma_2\sin\gamma_3|2\rangle\nonumber\\
&&+\sin\gamma_1\sin\gamma_2\cos\gamma_3|3\rangle,\nonumber\\
&U(\gamma_1,\gamma_2,\gamma_3)|2\rangle=|\phi_2\rangle=&-\sin\gamma_1\sin\gamma_2\cos\gamma_3|0\rangle\nonumber\\
&&+\sin\gamma_1\sin\gamma_2\sin\gamma_3|1\rangle\nonumber\\
&&+\cos\gamma_1|2\rangle-\sin\gamma_1\cos\gamma_2|3\rangle,\nonumber\\
&U(\gamma_1,\gamma_2,\gamma_3)|3\rangle=|\phi_3\rangle=&\sin\gamma_1\sin\gamma_2\sin\gamma_3|0\rangle\nonumber\\
&&+\sin\gamma_1\sin\gamma_2\cos\gamma_3|1\rangle\nonumber\\
&&-\sin\gamma_1\cos\gamma_2|2\rangle-\cos\gamma_1|3\rangle.\nonumber\\
\end{eqnarray}
Then we have
\begin{eqnarray}
&|\Phi\rangle_{ABC}=&\frac {1}{2}(|\phi_0\rangle_{A}|\Psi_0\rangle_{BC}+|\phi_1\rangle_{A}|\Psi_1\rangle_{BC}\nonumber\\
&&+|\phi_2\rangle_{A}|\Psi_2\rangle_{BC}+|\phi_3\rangle_{A}|\Psi_3\rangle_{BC}),
\end{eqnarray}
where
\begin{eqnarray}
&|\Psi_0\rangle_{BC}=&\cos\gamma_1|00\rangle+\sin\gamma_1\cos\gamma_2|11\rangle\nonumber\\
&&+\sin\gamma_1\sin\gamma_2\cos\gamma_3|22\rangle\nonumber\\
&&+\sin\gamma_1\sin\gamma_2\sin\gamma_3|33\rangle,\nonumber\\
&|\Psi_1\rangle_{BC}=&-\sin\gamma_1\cos\gamma_2|00\rangle+\cos\gamma_1|11\rangle\nonumber\\
&&-\sin\gamma_1\sin\gamma_2\sin\gamma_3|22\rangle\nonumber\\
&&+\sin\gamma_1\sin\gamma_2\cos\gamma_3|33\rangle,\nonumber\\
&|\Psi_2\rangle_{BC}=&-\sin\gamma_1\sin\gamma_2\cos\gamma_3|00\rangle\nonumber\\
&&+\sin\gamma_1\sin\gamma_2\sin\gamma_3|11\rangle\nonumber\\
&&+\cos\gamma_1|22\rangle-\sin\gamma_1\cos\gamma_2|33\rangle,\nonumber\\
&|\Psi_3\rangle_{BC}=&\sin\gamma_1\sin\gamma_2\sin\gamma_3|00\rangle\nonumber\\
&&+\sin\gamma_1\sin\gamma_2\cos\gamma_3|11\rangle\nonumber\\
&&-\sin\gamma_1\cos\gamma_2|22\rangle-\cos\gamma_1|33\rangle.\nonumber\\
\end{eqnarray}

After Alice measures particle $A$, the initial state will be
projected onto the measurement basis vectors with the appropriate
probability. She has to convey to Bob by classical communication
whether to apply the corresponding unitary transformation
\begin{eqnarray}
U_1=\left(\begin{array}{cccc} 0&1&0&0\\
-1&0&0&0\\
0&0&0&1\\0&0&-1&0\\\end{array}\right),\nonumber\\
U_2=\left(\begin{array}{cccc} 0&0&1&0\\
0&0&0&-1\\
-1&0&0&0\\0&1&0&0\\\end{array}\right),\nonumber\\
U_3=\left(\begin{array}{cccc} 0&0&0&-1\\
0&0&-1&0\\
0&1&0&0\\1&0&0&0\\\end{array}\right)\end{eqnarray}  on his particle
$B$ or do nothing. It means that  Alice's measurement outcomes
$|\phi_0\rangle$, $|\phi_1\rangle$,  $|\phi_2\rangle$, and
$|\phi_3\rangle$ correspond to unitary transformations $I$, $U_1$,
$U_2$, and $U_3$, respectively. Here $I$ is the identity operator.

 Next  Bob
constructs a  measurement basis and performs another projective
measurement on particle $B$, the relation between the measurement
basis  $\{\eta_0,\eta_1,\eta_2,\eta_3\}$ and the computational basis
$\{|0\rangle, |1\rangle, |2\rangle, |3\rangle$
 is  given by
\begin{eqnarray}
&&|\eta_0\rangle=\frac {1}{2}(|0\rangle+e^{-i\alpha_1}|1\rangle+e^{-i\alpha_2}|2\rangle+e^{-i\alpha_3}|3\rangle),\nonumber\\
&&|\eta_1\rangle=\frac {1}{2}(|0\rangle+ie^{-i\alpha_1}|1\rangle-e^{-i\alpha_2}|2\rangle-ie^{-i\alpha_3}|3\rangle),\nonumber\\
&&|\eta_2\rangle=\frac {1}{2}(|0\rangle-ie^{-i\alpha_1}|1\rangle-e^{-i\alpha_2}|2\rangle+ie^{-i\alpha_3}|3\rangle),\nonumber\\
&&|\eta_3\rangle=\frac
{1}{2}(|0\rangle-e^{-i\alpha_1}|1\rangle+e^{-i\alpha_2}|2\rangle-e^{-i\alpha_3}|3\rangle).
\end{eqnarray}
After Bob measures particle $B$, he will inform  Charlie of his
measurement result via a classical communication. Charlie can employ
an appropriate unitary operation to convert it to the prepared state
$|\psi\rangle$. For example, if Alice's measurement result is
$|\phi_1\rangle_A$, the state of particle $B$ and $C$, as shown in
Eqs. (15) and (16), will collapse into $|\Psi_1\rangle_{BC}$. After
Bob receives Alice's measurement result $|\phi_1\rangle_{A}$, he
first carries out the  unitary transformation $U_1$ described in
Eq.(17) on particle $B$. That is, the unitary operation $U_1$ will
transform the state $|\Psi_1\rangle_{BC}$ into
\begin{eqnarray}
&|\Psi'_1\rangle=&\sin\gamma_1\cos\gamma_2|10\rangle+\cos\gamma_1|01\rangle\nonumber\\
&&+\sin\gamma_1\sin\gamma_2\sin\gamma_3|32\rangle+\sin\gamma_1\sin\gamma_2\cos\gamma_3|23\rangle.\nonumber\\
\end{eqnarray}
Next, Bob performs the projective measurement on particle $B$ in the
basis described in Eq.(18). According to Bob's different measurement
result $|\eta_i\rangle$, Charlie needs to perform the corresponding
unitary operation $U_i(C)$ on particle $C$, $U_i(C)$ may take the
form of the following $4\times 4$ matrix
\begin{eqnarray}
&&U_0(C)=\left(\begin{array}{cccc} 0&1&0&0\\
1&0&0&0\\
0&0&0&1\\0&0&1&0\\\end{array}\right),\nonumber\\
&&U_1(C)=\left(\begin{array}{cccc} 0&1&0&0\\
i&0&0&0\\
0&0&0&-1\\0&0&-i&0\\\end{array}\right),\nonumber\\
&&U_2(C)=\left(\begin{array}{cccc} 0&1&0&0\\
-i&0&0&0\\
0&0&0&-1\\0&0&i&0\\\end{array}\right),\nonumber\\
&&U_3(C)=\left(\begin{array}{cccc} 0&1&0&0\\
-1&0&0&0\\
0&0&0&1\\0&0&-1&0\\\end{array}\right).\end{eqnarray}

The RSP is completed. Similarly, for other collapsed state
corresponding to Alice's measurement result, Bob can employ an
appropriate unitary operation in Eq.(17) or do nothing and perform
the projective measurement on particle $B$ in the basis described in
Eq.(18). Here we do not depict them one by one anymore. As a
summary, Bob's measurement outcomes corresponding to Alice's other
measurement results, and Charlie's corresponding unitary operations
to Bob's measurement results are listed in Table II.

\begin{table}
\caption{Alice's measurement outcome for particle $A$ (AMO), Bob's
measurement outcome for particle $B$ (BMO), and Charlie's
appropriate unitary operation (CAUO)} {
\begin{tabular}{|c|c|c|} \hline
 AMO & BMO & CAUO \\\hline
$|\phi_0\rangle_A$ & $|\eta_0\rangle_B$ & $I$ \\\hline
$|\phi_0\rangle_A$ & $|\eta_1\rangle_B$ & ${\rm diag}
(1,i,-1,-i)$\\\hline
 $|\phi_0\rangle_A$ & $|\eta_2\rangle_B$ & ${\rm
diag} (1,-i,-1,i)$
\\\hline
$|\phi_0\rangle_A$ & $|\eta_3\rangle_B$ & ${\rm diag} (1,-1,1,-1)$
\\\hline
$|\phi_2\rangle_A$ & $|\eta_0\rangle_B$ & $\left
(\begin{array}{cc}0& A_1\\A_1& 0\end{array}\right), A_1={\rm
diag}(1,1)$
\\\hline $|\phi_2\rangle_A$ & $|\eta_1\rangle_B$ & $ \left
(\begin{array}{cc}0& A_2\\-A_2& 0\end{array}\right), A_2={\rm
diag}(1,i )$
\\\hline
$|\phi_2\rangle_A$ & $|\eta_2\rangle_B$ &  $\left
(\begin{array}{cc}0& A_3\\-A_3& 0\end{array}\right), A_3={\rm
diag}(1,-i)$
\\\hline
$|\phi_2\rangle_A$ & $|\eta_3\rangle_B$ &  $\left
(\begin{array}{cc}0& A_4\\A_4& 0\end{array}\right), A_4={\rm
diag}(1,-1)$ \\\hline $|\phi_3\rangle_A$ & $|\eta_0\rangle_B$ &
$\left (\begin{array}{cc}0& A_5\\A_5& 0\end{array}\right),
A_5=\left(\begin{array}{cc}0&1\\1&0\end{array}\right)$
\\\hline $|\phi_3\rangle_A$ & $|\eta_1\rangle_B$ & $\left
(\begin{array}{cc}0& A_6\\-A_6& 0\end{array}\right),
A_6=\left(\begin{array}{cc}0&1\\i&0\end{array}\right)$
\\\hline $|\phi_3\rangle_A$ & $|\eta_2\rangle_B$ & $\left
(\begin{array}{cc}0& A_7\\-A_7& 0\end{array}\right),
A_7=\left(\begin{array}{cc}0&1\\-i&0\end{array}\right)$
\\\hline $|\phi_3\rangle_A$ & $|\eta_3\rangle_B$ &  $\left
(\begin{array}{cc}0& A_8\\A_8& 0\end{array}\right),
A_8=\left(\begin{array}{cc}0&1\\-1&0\end{array}\right)$\\\hline
\end{tabular}}
\end{table}

By the above analysis, we may conclude that the essence of our
protocol is first preparing a point of the polar circle, and then
adding the information of the equatorial state. Now, we suppose that
Alice and Bob want to remotely prepare a known $d$-level quantum
state at Charlie's location. However, not all the qudits can be
remotely prepared according to Ref. [6], in which the authors have
shown that the qudit in real Hilbert space can be remotely prepared
when the dimension is 2, 4, or 8. So here we let  $d=8$. The state
of an eight-dimensional system can be written as
\begin{equation}
|\psi\rangle=\sum^7_{i=0}\cos\theta_ie^{i\varphi_i}|i\rangle, ~~
\sum_{i=0}^7|\cos\theta_i|^2=1.
\end{equation}
 Without loss of generality, we set $\varphi_0=0$. According to the
 analogous procedure described above, the corresponding qudit to be
 prepared can be remotely prepared exactly onto the particle at
 Charlie's location. The measurement basis chosen by Alice can be
 obtained by $V_i|\psi\rangle$. The unitary operation $V_i$ needed
 is the same as those for eight-dimensional RSP in Ref. [6]. Bob's
 measurement basis is written as $\{|\eta_j\rangle
 =\sum^7_{k=0}e^{(\pi i/4)jk}e^{i\varphi_j}|k\rangle\}^7_{j=0}
 .$

\section {Conclusions}

In summary,  we have presented a two-step protocol for the exact
remote state preparation of an arbitrary qubit using one
three-particle GHZ state as the quantum channel. Only a
single-particle von Neumann measurement and local operation are
necessary. It has been shown that the overall information of the
qubit, can be divided into two different parts, which are expressed
by $\theta$ and $\varphi$ respectively. We must first prepare the
part $\theta$ and then prepare the remainder part $\varphi$, which
can not be transposed. This indicates that the two parts of
information are not equal with each other. Generalization of this
protocol for higher-dimensional
 Hilbert space systems among three parties is also presented. Moreover, it should be noticed that in this protocol, the
 information $\theta$ and $\varphi$  may be at different locations. So this protocol may be useful in the quantum information field,
  such as quantum state sharing, converging the split information at one point, etc. We hope this will provide new insight for
   investigating more extensive quantum information processing procedures.\\

{\noindent\bf Acknowledgments}\\[0.2cm]

 This work was supported by the National Natural Science Foundation of China under Grant No: 10671054, Hebei Natural Science Foundation of China under Grant
No: 07M006, and  the Key Project of Science and Technology Research
of Education Ministry of China under Grant No: 207011.

\end{document}